\documentclass{iau}

\usepackage{epsfig}

\newfont{\myfont}{cmmib10}

\newcommand{\btheta}{\hbox{\myfont \symbol{18} }}

\title[IAUS291.~~Interstellar scattering] 
{Interstellar scattering --- New diagnostics of pulsars and the ISM} 

\author[J.-P. Macquart]  
{Jean-Pierre Macquart
}

\affiliation{ICRAR/Curtin Institute of Radio Astronomy, Curtin
  University, Perth WA 6845, Australia \\ [\affilskip] 
ARC Centre of Excellence for All-Sky Astrophysics (CAASTRO)
\\ [\affilskip] 
email:
{\tt J.Macquart@curtin.edu.au} \\  
}

\pubyear{2012}
\volume{291}  
\jname{\mbox{Neutron Stars and Pulsars: Challenges and Opportunities after 80 years}}
\editors{J. van Leeuwen, ed.} 
\begin{document}

\maketitle

\begin{abstract}
Extreme Scattering Events and pulsar secondary spectra have highlighted fundamental problems in our understanding of the dynamics of interstellar turbulence.  We describe some of these problems in detail and present the theory behind the technique of speckle imaging, which offers a prospect of revealing fundamental properties of the turbulence.  It also offers the prospect of resolving pulsar magnetospheres on $\sim 10\,$nas scales. 
\end{abstract}


\firstsection 
\section{Why bother about turbulence?}

Compilations of pulsar scattering observations reveal that, in some average sense, the power spectrum of electron density fluctuations in the ISM follows a power law on scales from $\sim 10^6\,$m up to $10^{14-18}\,$m, with a power-law index close to $11/3$, the value expected from Kolmogorov turbulence (Armstrong, Rickett \& Spangler 1995).  In the Kolmogorov view of turbulence, mechanical and magnetic energy from macroscopic processes, such as stellar winds and explosions, cascades to smaller scales via a succession of self-similar kinetic-energy conserving turbulent eddies.  It is surprising that the predictions of Kolmogorov theory, which applies to incompressible hydrodynamic turbulence, should resemble in any way the structure of the interstellar medium. 

The ISM is not incompressible over a large range of scales of interest (Luo \& Melrose 2006), and is intrinsically {\it magneto}-hydrodynamic.  Indeed, the magnetic field must play a strong role in mediating the interstellar turbulent cascade because the presence of diffractive scintillation in pulsars shows that turbulence persists on scales many orders of magnitude below the collisional mean free paths of electrons and protons (Lithwick \& Goldreich 2001).   Moreover, there is no consensus on the most basic physics of how energy is mediated between large and small scales in the highly magnetized ISM, nor is there agreement on the interrelation between the turbulent velocity, magnetic field and density fluctuations (Elmegreen \& Scalo 2004, \S4.12-13).

Several key observations over the last decade have added a number of complications to our view of interstellar turbulence.  The most notable has been the realization that there exist local pockets of anomalously strong turbulence.  This is evident in the intermittency of scintillations of intra-day variable quasars (Lovell et al.\,2008).  It is also manifest in the anomalous turbulence associated with tiny ($\sim$10$^{11}\,$m) ionized clouds that are prevalent throughout the ISM and that are responsible for Extreme Scattering Events (Fielder et al.\,1987).  The discovery of strong parabolic arcs in the secondary spectra of many pulsars has also revealed that in many instances the scattering is often highly localized along the line of sight, and that the turbulence itself appears to be highly anisotropic.
Attempts to incorporate these anomalous properties into a physical framework of the ISM have proven problematic and controversial (Spangler \& Vazquez-Semadeni 2007).

We elucidate the problems exposed by ESEs and pulsar secondary spectra in \S2, while \S3 describes the method of speckle imaging that is allowing us to directly ``image'' these scattering structures in the ISM.  In \S4 we present some preliminary limits on the nano-arcsecond structure of the pulsars whose radiation is being scattered by some of these anomalous scattering structures.  The final section describes the prospects for solving some of the fundamental questions that relate to interstellar turbulence.



\section{ESEs, Parabolic Arcs and Anisotropic scattering}

Extreme Scattering Events (ESEs) are intensity excursions exhibited by some compact quasars and pulsars and lasting between 10 and 50\,days (Fiedler et al.\,1987; Romani et al.\,1987). The symmetric nature of their lightcurves argues that they are due to the passage of 4-70$\times$10$^{10}\,$m sized cloud-like features across the lines of sight.  The observed event rate of 0.013 source$^{-1}$\,year$^{-1}$ means that the clouds are common, with an estimated volume density of one per $\sim 10^{-5}\,$pc$^3$ (Fiedler et al. 1994; Walker \& Wardle 1998). 
  
The existence of ESEs is problematic because their observed optical properties imply, at face value, internal pressures that exceed the diffuse ISM's by three orders of magnitude, assuming temperatures comparable to the diffuse warm ISM (Spangler \& Vazquez-Semadeni 2007). In a simple model in which a plasma overdensity refracts the radio waves sufficiently to reproduce the caustic peaks observed in ESE lightcurves, column densities of $\sim 10^{19}\,$cm$^{-2}$ are required.  The volume density depends on the elongation of the structure along the line of sight; for an elongation $\eta = 100 \eta_2$ the density is $\sim 10^3\,\eta_2^{-1}\,$cm$^{-3}$ (Romani et al.\,1987).  However, other models may explain ESEs without recourse to such extreme properties; Pen \& King (2012) have recently proposed that ESE may instead be interpreted in terms of {\it under}dense sheets in the ISM.

There is evidence from pulsar scattering that much of the turbulence in the ISM is highly localised and anisotropic.  One of the principal means of gleaning this information from pulsar scattering measurements is via the secondary spectrum.  The secondary spectrum, $A(\tau,\omega)$, is the squared amplitude of the two dimensional Fourier transform of the dynamic spectrum of a pulsar's intensity scintillations, $I(\nu,t)$.  In the secondary spectrum, the conjugate of observing frequency is the delay, while the Fourier conjugate of time is Doppler frequency.  In the regime of strong scattering from a thin scattering screen, one can represent the received wavefield as the sum of wavefields from a set of stationary phase points (or speckles) on the surface of the scattering disk, $u = \sum_j a_j e^{i \Phi_j}$,  
with each stationary phase point possessing an amplitude $a_j$ and phase $\Phi_j = \phi({\bf x}_j) + ({\bf x}_j-\beta {\bf r})^2/2 r_{\rm F}^2$, where $\phi({\bf x}_j)$ is the phase delay imposed by the scattering medium at the position, ${\bf x}_j$, of the stationary phase point, and ${\bf r}$ is the location of the telescope on the observer's plane.  The distance to the scattering screen, $D_s$, and the distance to the pulsar, $D_p$, also effect the total phase delay via the Fresnel scale, $r_{\rm F} = ( \beta D_s /k)^{1/2}$, where $\beta=1-D_s/D_p$.  The speckles emanate from positions $\btheta_j = ({\bf x}_j -\beta {\bf r})/D_s$ on the scattering disk.

The intensity scintillation pattern, $I(\nu,t) = u u^*$, is the result of the interference of each speckle with every other speckle on the scattering disk, and the resulting secondary spectrum takes the form (e.g. Walker et al. 2004),
\begin{eqnarray}
A(\tau,\omega) &\propto& \sum_{j,k} a_j a_k \left[ \delta (\tau-\tau_{jk}) \delta(\omega - \omega_{jk}) + \delta (\tau+\tau_{jk}) \delta (\omega+\omega_{jk}) \right], \label{SSeq} \\ 
\hbox{where} &\tau_{jk}& = \frac{D_s}{2\,c\,\beta} (\theta_j^2 - \theta_k^2) + \left[ \frac{\phi_j}{2 \pi \nu} - \frac{\phi_k}{2 \pi \nu} \right],  \quad \omega_{jk} = \frac{1}{\lambda} \left( \btheta_j - \btheta_k \right) \cdot {\bf v}_{\rm eff}. \label{omegaEq} 
\end{eqnarray}
  
The location of power in the secondary spectrum is predominately dictated by the positions of the speckles\footnote{The terms in square brackets in eq.\,(\ref{omegaEq}) make only a small contribution to the overall delay.}.  For any given pair of speckles, $j$ and $k$, power appears at both the co-ordinates $(\tau_{jk},\omega_{jk})$ and $(-\tau_{jk},-\omega_{jk})$.  This symmetry is a reflection of the fact that $A(\tau,\omega)$ is derived from the Fourier transform of a real quantity, namely $I(\nu,t)$.  The effective scintillation velocity, ${\bf v}_{\rm eff}$ also influences the Doppler frequency of the speckles; it is usually dominated by the pulsar velocity, ${\bf v}_p$, but it may in principle also be affected by the peculiar velocity of the screen, ${\bf v}_s$ or of the Earth, ${\bf v}_\oplus$: ${\bf v}_{\rm eff} = \beta^{-1} {\bf v}_{\rm ISS} = \beta^{-1} \left[ (1-\beta) {\bf v}_p + \beta {\bf v}_\oplus - {\bf v}_{\rm s} \right].$

\begin{figure}
\psfig{file=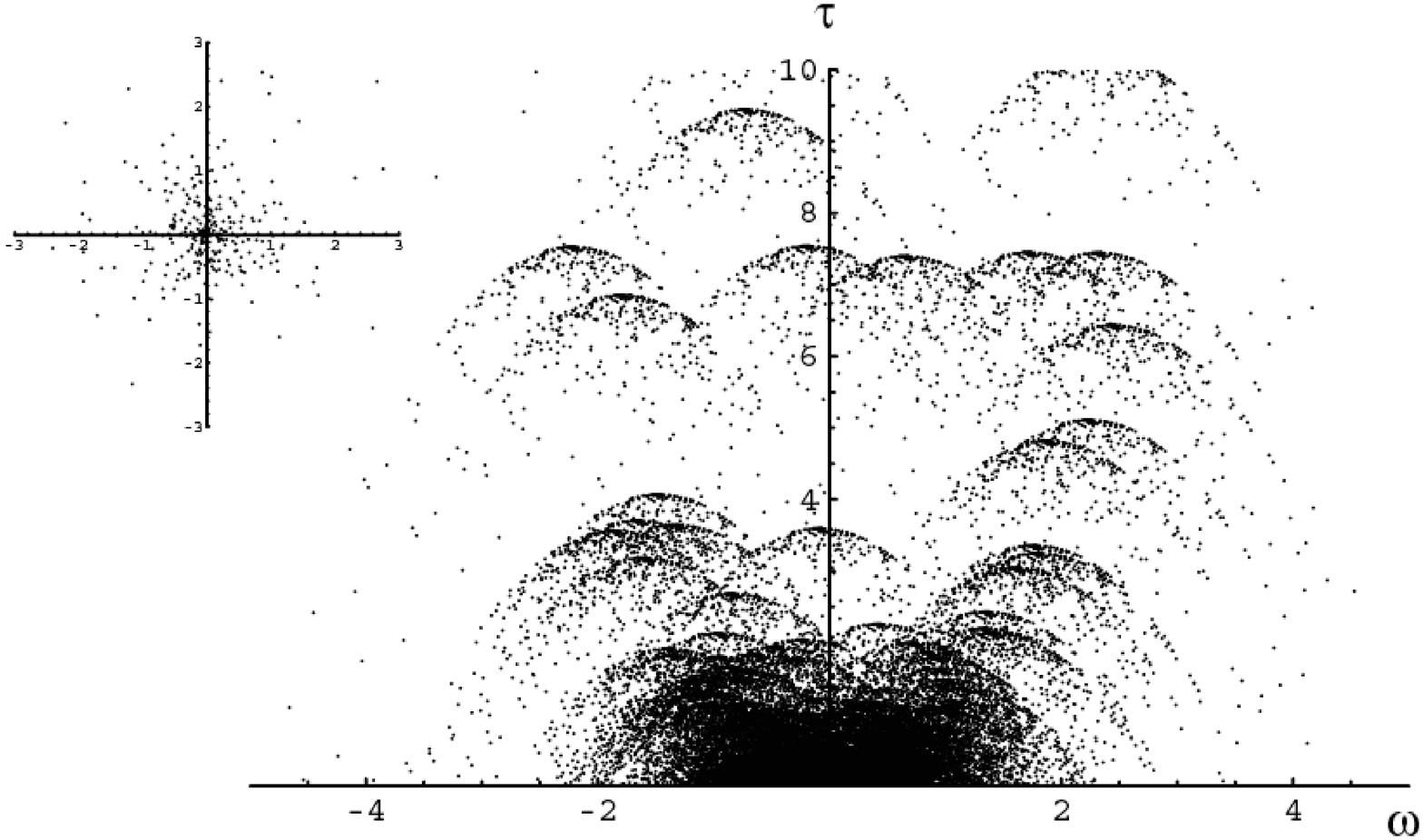, scale=0.21} 
\psfig{file=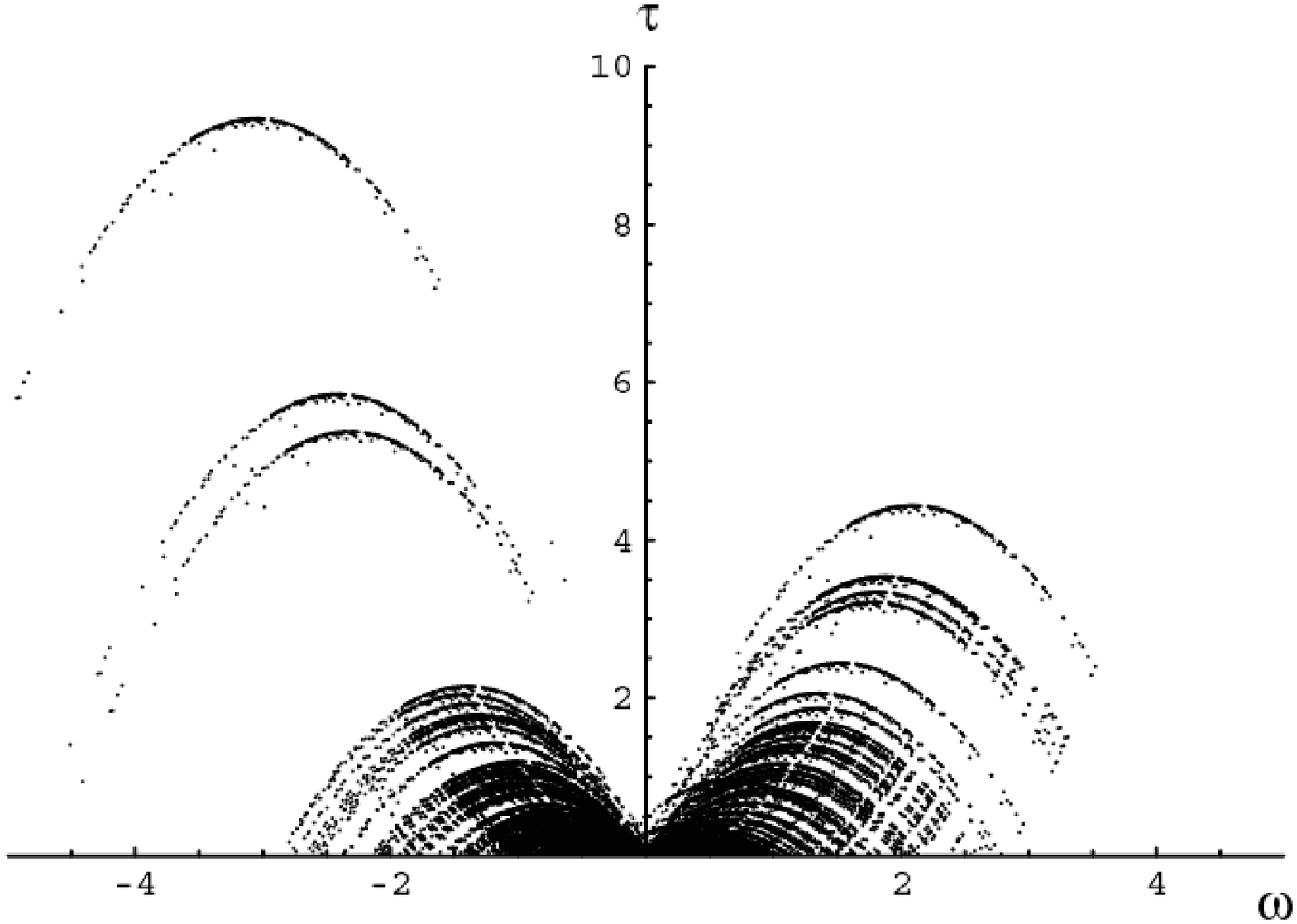, scale=0.21}
\caption{Left: the secondary spectrum corresponding to 300 speckle points distributed according to  isotropic Kolmogorov turbulence.  The inset panel shows the spatial speckle distribution.  Right: the same as the left figure except that the distribution of points has been made anisotropic by reducing the $\theta_y$ co-ordinates of the speckles by a factor of $5$.} \label{fig:SS}
\end{figure}

The prescription given by eqs.\,(\ref{SSeq})-(\ref{omegaEq}) affords a geometric interpretation of the secondary spectrum.  Consider interference caused by a highly elongated speckle pattern, in which all the speckles make a constant angle, $\alpha$, to the scintillation velocity.  Interference between a speckle at location $\btheta_j$ with the bright ``core'' of the image at $\btheta_k=0$ traces out a parabola with the locus $(\tau_0,\omega_0) = (D_s \theta^2/2c\beta,\theta_j v_{\rm eff} \cos \alpha/\lambda)$.  Allowing variation in $\btheta_k \neq 0$, we obtain the locus of an inverted parabolae whose apex occurs at co-ordinates $(\tau_0, \omega_0)$.  If, instead, the distribution of speckles is not highly elongated, variations in $\alpha$ causes power to lie interior to a bounding parabola, as shown in Figure \ref{fig:SS}.  

Many pulsars exhibit strong, sharply defined parabolic arcs of the sort seen in the right of Figure \ref{fig:SS}.   This reveals two important qualities of the scattering.  1. The scattering is highly anisotropic.  If it were otherwise, power in the secondary spectrum would instead lie interior to the main parabola.  2.  The scattering occurs in localized patches, on a scattering screen that is thin, a few percent of the total pulsar distance.  Variation in $D_s$ and $\beta$ would otherwise smear out the locations of points in the secondary spectrum.  

Unfortunately, eqs.\,(\ref{SSeq})-(\ref{omegaEq}) do not enable an unambiguous reconstruction of the speckle distribution because the angle $\alpha$ is not known for any given speckle.  The delay only measures $\theta_j^2$ and the Doppler frequency only measures $\theta_j v_{\rm eff} \cos \alpha/\lambda$. This degeneracy can be broken by instead measuring the scintillations in the interferometric visibility on intercontinental baselines, as we now describe.

\section{Speckle Imaging}

It is possible to form a complete image of the distribution of speckles on the scattering disk by forming the secondary spectrum of the interferometric visibility.   As we show below, the distribution of power in the visibility secondary spectrum is very close to that observed in the intensity secondary spectrum, but the visibilities provide information on the astrometric phase shift associated with each pair of speckles.  In essence, one is able to image the disk by using the secondary spectrum to isolate the wavefield of each pair of interfering speckles, and then measure the phase of this isolated wavefield from the visibility  secondary spectrum to localise the speckle positions to extremely high precision.

To see how this works, consider the visibility that a two-element interferometer would measure from a scattered pulsar.  Quantities measured with the first and second receiving elements are labelled with subscripts $1$ and $2$ respectively.  The elements are placed at locations ${\bf r}_1= {\bf R} - \Delta {\bf r}/2$ and ${\bf r}_2= {\bf R} + \Delta {\bf r}/2$ on the observer's plane, with ${\bf R}$ the mean position of the telescopes, and $\Delta {\bf r}$ their relative displacement.  The measured visibility is, 
\begin{eqnarray}
V(\nu,t) = u_1 u_2^* = \sum_{j,k}^N a_{1,j} a_{2,k} [\cos(\Phi_{1,j} - \Phi_{2,k}) + i \sin (\Phi_{1,j} - \Phi_{2,k}) ],
\label{Vis}
\end{eqnarray} 
where $\Phi_{jk} = \Phi_{1,j} - \Phi_{2,k}$ is the phase difference between the $j$th stationary phase point measured at station 1 and the $k$th stationary phase point measured at station 2.
In the limit in which the bandwidth and observing duration are large, the Fourier-transform of the visibility dynamic spectrum reduces to,
\begin{eqnarray}
\tilde V (\tau,\omega) &=& \frac{1}{2 \pi} \sum_{j,k}^N a_{1,j} a_{2,k} \Big\{ 
\exp[i \Phi_{jk}^0] \delta (\tau + \tau_{jk}) \delta (\omega + \omega_{jk})  \Big\}, \\
\omega_{jk} &=& \frac{1}{2\pi} \frac{\partial (\Phi_{1,j}-\Phi_{2,k})}{\partial t} 
=  \frac{1}{\lambda \beta} (\btheta_{1,j} - \btheta_{2,k}) \cdot {\bf v}_{\rm eff} + 
			\frac{\beta}{\lambda D_s}  \Delta {\bf r} \cdot {\bf v}_{\rm eff},  \\
 \tau_{jk} &=& \frac{1}{2 \pi} \frac{\partial (\Phi_{1,j}-\Phi_{2,k})}{\partial \nu}
=  \frac{D_s (\theta_{1,j}^2 - \theta_{2,k}^2 )}{2 c \beta} + (\btheta_{1,j} + \btheta_{2,k}) \cdot \frac{\Delta {\bf r}}{c} - \frac{\phi_{1,j}}{2\pi \nu} + \frac{\phi_{2,k}}{2 \pi \nu} , \\
\Phi_{jk}^0 
&=& \phi_j - \phi_k + \frac{D_s^2}{2 r_{\rm F}^2} \left( \btheta_{1,j}^2 - \btheta_{2,k}^2 \right) 
- \frac{\beta D_s}{2 r_{\rm F}^2} ( \btheta_{1,j} + \btheta_{2,k}) \cdot \Delta {\bf r},
\end{eqnarray}
where we write $\btheta_{1,j} = ({\bf x}_{j} - \beta {\bf R})/D_s$, $\btheta_k = ({\bf x}_k - \beta {\bf R})/D_s$.  Unlike its single-dish counterpart, the visibility secondary spectrum is not symmetric under the operation $(\tau,\omega) \rightarrow (-\tau,-\omega)$.  This is because $\omega_{jk} \neq -\omega_{kj}$ and $\tau_{jk}  \neq - \tau_{kj}$.  However, if the visibilities are measured on a baseline small compared to the scale of the scintillation pattern (i.e. $\Delta {\bf r} \ll D_s \btheta_j$), the {\it amplitude} of the visibility secondary spectrum is effectively identical to that observed in the intensity secondary spectrum.  

An important difference between the intensity and visibility secondary spectrum is that the latter contains an $\exp[i \Phi^0_{jk}]$ phase term that is no longer antisymmetric.  This is because the phase of each speckle measured at two widely-separated telescopes differs slightly because of astrometric phase term. In terms of the formalism introduced here, we see that when $\Delta {\bf r} =0$, the term proportional to the $(\btheta_j + \btheta_k) \cdot \Delta {\bf r}$ destroys the odd symmetry $\Phi^0_{jk} = - \Phi^0_{kj}$.   If one adds the contribution from the $j,k$ and $k,j$ terms in the secondary spectrum, it is possible to isolate the astrometric phase of each pair of speckles and determine the projection of the position of each pair of speckles, $\btheta_j + \btheta_k$ along the baseline $\Delta {\bf r}$.  Thus, measurements along two baselines are sufficient to to determine the position of each pair of speckles.  This is the basis of scintillation speckle imaging.

This technique was first applied to PSR B0834$+$06 by Brisken et al.\,(2010).  These 327\,MHz observations revealed a number of intriguing properties of the scattering medium:
\begin{itemize}
\item The scattering disk was composed of two separate structures, separated by 9\,AU.
\item Speckles along the primary scattering disk, which was 16\,AU long, were distributed anisotropically, with the ratio of the major to minor axis of the disk being at least 27:1.  The distribution of speckles along the long axis of the primary scattering disk does not resemble that expected of Kolmogorov turbulence.
\item The secondary scattering disk contributed about 4\% of the total power.  It is tempting to speculate on the origin of this feature.  One possibility, given the strong scattering properties the feature must possess in order to scatter a substantial amount of off-axis power back into the line of sight, is that it may be a cloud of the sort that is implicated in Extreme Scattering Events.  This hypothesis, however, is difficult to test because ESEs are characterised by their optical properties when viewed on-axis with respect to a background source, whereas this object was off-axis by $>20\,$mas.
\end{itemize}

\subsection{Magnetic field limits}
Given the highly anisotropic scattering inferred towards PSR B0834$+$06 and other pulsars which display strong parabolic arcs, it seems clear that the magnetic field plays an important role in the turbulent dynamics.  One means of probing the magnetic field is to search for small rotation measure (RM) fluctuations associated with the scattering medium.  If RM fluctuations are present, the left and right-hand circularly polarized components of the pulsar radiation will experience slightly different phase delays, and this will cause different scintillations in each sense of circular polarization (Macquart \& Melrose 2000), which would be visible in the circular polarization secondary spectrum.  

Brisken et al. (2010) report that, for PSR B0834$+$06, no detectable scintillating circular polarization signal was detected at the 0.1\% level, and this places a limit on the RM difference of less than $1.2 \times 10^{-3}\,$rad\,m$^{-2}$ across AU scales on the scattering disk.   

\section{ISS as probes of nano-arcsecond pulsar structure}
The fact that speckle images exhibit structure across baselines of $>10\,$AU implies stringent constraints on the size of the pulsar emission region.  Interference between the primary and secondary scattering disks in the case of PSR B0834$+$06 means that the radiation from the pulsar is at least partially coherent on a baseline of $\delta=1.3 \times 10^{11}\,$m as viewed at the scattering screen.  This translates to a physical scale at the pulsar of $\delta (D_s - D_p) = 4700\,$km.  If the pulsar radiation had a gaussian angular brightness profile, the HWHM of the brightness distribution would be $\sim 850\,$km.

\begin{figure}
\centerline{\psfig{file=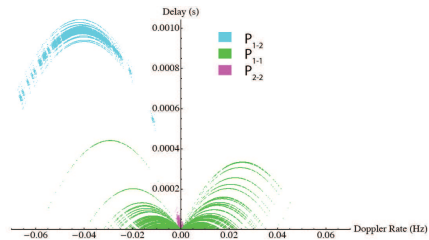, width=58mm}}
\caption{A schematic of the secondary spectrum of PSR 0834$+$06 found by Brisken et al.\,(2010).   The green points represent interference between speckles on the primary scattering disk, cyan represents interference between speckle pairs on the primary disk with those on the secondary scattering disk, and the purple points represent interference between adjacent speckles on the secondary scattering disk.} \label{fig:SSdist}
\end{figure}

Although the radiation must be at least partially coherent on 9\,AU baselines, it is difficult to determine the degree of coherence.  This is because one does not know what fraction of the total power should be received from the secondary scattering disk if the pulsar radiation were 100\% spatially coherent on this baseline.  One can in principle determine this by comparing the power associated with the interference between pairs of speckles on the primary disk, $P_{1-1}$, and pairs of speckles on the secondary scattering disk, $P_{2-2}$, with the power associated with interference between speckles on the primary disk with those on the secondary disk, $P_{1-2}$.  The pulsar is resolved if the quantity, ${\cal R} = {P_{1-1} P_{2-2}}/P_{1-2}^2$ significantly exceeds one.  Figure \ref{fig:SSdist} shows that $P_{1-1}$, $P_{2-2}$ and $P_{1-2}$ can all be measured from the secondary spectrum.  However, a complication arises because these three power quantities can be difficult to measure in practice, and it is difficult to relate ${\cal R}$ to a specific measurement of the pulsar angular size.

\section{The Future}
There are two obvious prospects for progress in this field at present.  The first, involving interstellar holography, which is not discussed in this short paper, has been advanced by the efforts of Walker et al. (2005, 2008).  Holography takes advantage of the highly redundant information provided in the dynamic spectrum about the speckle distribution: for $N$ speckles there are $N(N-1)/2$ interfering pairs measured in the secondary spectrum.  The concept of interstellar holography may be viewed as a deconvolution problem.  The Fourier transform of the intensity $I(\nu,t) = u(\nu,t) u^*(\nu,t)$, is just the autoconvolution of the Fourier-transformed wavefield: $\tilde I (\tau,\omega) = \tilde u(\tau, \omega) \star \tilde u^*(\tau,\omega)$.  

Recent holographic work performed by Ue-Li Pen and collaborators on PSR B0834$+$06 data has claimed a measurement of the pulsar's reflex motion (Pen et al. in prep.).  Holography was used to effectively descatter the pulsar radiation and permit extremely high S/N measurements of the pulsar's radiation.  By performing this holography over a succession of bins in pulse phase, it has been possible to measure a small but significant phase shift associated with the pulsar reflex motion.

A second prospect is related to measuring the motions of speckles in scattered pulsar images using a succession of speckle images over a period of weeks to months.  In most cases, the pulsar proper motion dominates the effective scintillation velocity, but ultra-high S/N astrometric imaging, of the sort performed on PSR B0834$+$06, offers the prospect of resolving motions of the individual speckle groups relative to the bulk motion.  This therefore offers the prospect of relating the density fluctuations associated with the scattering disk (which are, to some limited extent, recoverable with holographic techniques) with the underlying turbulent velocity fluctuations. 



\end{document}